# Analyzing Effects of Seasonal Variations in Wind Generation and Load on Voltage Profiles

Malhar Padhee and Anamitra Pal
School of Electrical, Computer, and Energy Engineering
Arizona State University
Tempe, AZ 85281, USA
Emails: {mpadhee, anamitra.pal}@asu.edu

Katelynn A. Vance
Virginia Electric & Power Company
d/b/a Dominion Virginia Power
Richmond, VA 23219, USA
Email: Katelynn.A.Vance@dom.com

*Abstract*—**This paper presents a methodology for building daily profiles of wind generation and load for different seasons to assess their impacts on voltage violations. The measurement-based wind models showed very high accuracy when validated against several years of actual wind power data. System load modeling was carried out by analyzing the seasonal trends that occur in residential, commercial, and industrial loads. When the proposed approach was implemented on the IEEE 118-bus system, it could identify violations in bus voltage profiles that the season-independent model could not capture. The results of the proposed approach are expected to provide better visualization of the problems that seasonal variations in wind power and load might cause to the electric power grid.**

## I. INTRODUCTION

Increased investment in fossil-free power generation is resulting in significant renewable energy penetration. The two most popular sources of renewable energy production are wind and solar. When their penetration percentage was small, these distributed energy resources (DERs) did not significantly affect the reliability of the bulk power system. However, due to the recent advances made in power electronics as well as the incentives provided by the Federal Government, it is expected that large numbers/sizes of DERs, especially wind, will be added to the power transmission network. Wind integration with the traditional grid will create new challenges that must be overcome before such generation schemes can be considered viable [1]. For instance, according to [2], there can be detrimental impacts on the stability of the power grid when the wind energy penetration becomes 20%–40% of the total power generation. Therefore, extensive research needs to be done to operate the power system with high reliability *and* high penetration percentage of DERs.

This paper focuses on scenarios where wind generation constitutes approximately half of the total generation. It is also assumed that wind is the only source of renewable energy present in the system. Many papers have been published in the past five years on wind energy integration. Ref. [3] analyzed the impact of doubly fed induction generator wind farms on power system transient stability by evaluating the transient energy margin index for different operating conditions. Ref. [4] proposed a novel framework to geometrically determine the impact of wind power uncertainty on the small-signal stability of bulk power systems. Ref. [5] minimized the hourly social cost in presence of wind generation by proposing a market-based probabilistic optimal power flow (OPF) that placed energy storage systems (ESSs) at key locations. Ref. [6] proposed a method for optimally allocating ESSs in a wind-integrated power system using a hybrid multi-objective particle swarm optimization (PSO) technique. Prior literature has often used the Weibull distribution to model annual wind speed, in which the wind speed-power curve relation is used to calculate the power output. However, due to the assumptions involved, such an approach may not accurately capture the seasonal variations that occur in wind power output. This paper proposes a measurement based approach for analyzing seasonal variations in wind generation. Load modeling has been usually done on an event-by-event basis [7], [8]. The effect of seasonal variations on the load profiles has also not been explored in great details yet. This paper models the system load by considering characteristics of different load types (residential, commercial, and industrial). To the best of the authors' knowledge, *this is the first paper that studies the simultaneous impact of seasonal variations in wind generation and system load on bus voltage profiles*. It is hypothesized that by using such a season-focused approach, the capability of previously proposed techniques (such as [5], [6]) can be further refined.

The wind power data for building the proposed measurement-based approach was obtained from the Bonneville Power Administration (BPA) website [9]. Wind power output for the BPA control area from 2007 to 2015 is recorded in MW for every half-hour. Using this data, wind power output models for a normative day for all four seasons were built. Monthly energy demand data for the years 2007 to 2015 for Oregon (where BPA is located) was obtained from the U.S. Energy Information Administration (EIA) [10], for different load types. Using this data, load models for a normative day for different seasons were built. The seasonal power demand was modeled on a half-hourly basis to match the corresponding wind power output. The wind power output and load models were then applied on a large test system to find the locations that are most vulnerable to seasonal voltage violations.

This work was partially supported by the U.S. Department of Energy (DOE) Grant DE-EE0007660.

The remainder of the paper is structured as follows. Section II presents the theoretical background for model and measurement-based wind power modeling as well as seasonal load modeling. Section III validates the measurement-based wind-modeling approach by splitting the BPA wind power data into training and testing datasets. Section IV presents the results that were obtained when the effects of season-independent and season-focused modeling approaches were compared. The IEEE-118 bus system was used for comparison. The concluding comments and future scope of work are presented in Section V.

## II. THEORETICAL BACKGROUND

### A. Model-based approach to wind power modeling

The Weibull distribution [11], [12] is conventionally used to statistically model the wind speed. The probability density function of this distribution is given by (1),

$$f_v(v) = \left(\frac{k}{\lambda}\right)\left(\frac{v}{\lambda}\right)^{k-1} \exp\left(-\left(\frac{v}{\lambda}\right)^k\right), 0 \leq v \leq \infty \quad (1)$$

where $v$ represents the wind speed at the present instant, $k$ represents the shape coefficient and $\lambda$ is the scale coefficient. The maximum likelihood estimates of the Weibull distribution parameters are computed using curve fitting for historic time-stamped wind speed data [5]. For the model-based approach, 30-minute average wind speed data was obtained from the BPA website [13]. The half-hourly wind speed values in miles per hour are converted to half-hourly wind power outputs by using the wind speed-power curve relation as shown in (2).

$$P_h = \begin{cases} 0, & WS_h \leq V_{ci} \text{ or } WS_h \geq V_{co} \\ Prw\left(\frac{WS_h - V_{ci}}{V_r - V_{ci}}\right), & V_{ci} < WS_h < V_r \\ Prw, & V_r \leq WS_h < V_{co} \end{cases} \quad (2)$$

In (2), $V_{ci}$ represents the cut-in speed in meter/sec, $V_{co}$ represents the cut-out speed in meter/sec, $V_r$ represents the rated speed in meter/sec and $Prw$ represents the rated wind farm output in MW. For each day of the year, the 48 half-hourly wind power output values are normalized with respect to the first half-hourly wind power output value of the same day. Finally, the normative wind power output, $Pn$ for the $h$th half hour of any day $d$ of the year is obtained in (3).

$$Pn_h = \frac{\sum_{d=1}^{365} P_{h+(d-1)\times 48}}{365} \quad (3)$$

In (3), $h = \{1, ..., 48\}$.

### B. Measurement-based approach to wind power modeling

In this approach, 30-minute wind power output data for the years 2007 to 2011 was obtained from the BPA website [9]. For each day, the 48 half-hourly wind power output values are normalized with respect to the half-hourly wind power output obtained at 12:30 AM on the same day. Then, for each of the five years, the normalized half-hourly wind power outputs are grouped into four seasons: winter (December-February), spring (March-May), summer (June-August) and fall (September-November). Later, for each season of the five years, a normative seasonal wind power output model is built. The mean wind power output of the $h$th half hour of any normative season is,

$$MPseason_h = \frac{\sum_{y=1}^{5} Pseason_{h,y}}{5} \quad (4)$$

where $Pseason_{h,y}$ represents the $h$th half-hourly wind power outputs of any season of year $y$. The $h$th half-hourly wind power output of the normative season is the half-hourly wind power output of the year $y$, $Pseason_{h,y}$ that minimizes $|Pseason_{h,y} - MPseason_h|$, and is denoted as $NPseason_h$. The minimum variance in wind power output during the $h$th half-hour of the normative season is the minimum value of $(Pseason_{h,y} - MPseason_h)$, and is denoted by $MINVNPseason_h$. The maximum variance in wind power output during the $h$th half-hour of the normative season is the maximum value of $(Pseason_{h,y} - MPseason_h)$, and is denoted by $MAXVNPseason_h$. For each season, a normative day wind power output model is then built, which is representative of the wind power variation for each day of the season. The mean wind power output of the $h$th half hour of any day of the season is given by (5).

$$MDPseason_h = \frac{\sum_{d=1}^{N} NPseason_{h+(d-1)\times 48}}{N} \quad (5)$$

In (5), $h = \{1, ..., 48\}$ and $N$ is the length of season, in days. The $h$th half-hourly wind power output of a normative day of a season, is the wind power output during the $h$th half-hour of day $d$ of the normative season, $NPseason_{h+(d-1)\times 48}$ which minimizes $|NPseason_{h+(d-1)\times 48} - MDPseason_h|$, and is denoted by $NORMPseason_h$. The minimum variance in wind power output during the $h$th half-hour of any day of the season is the minimum value of $(NPseason_{h+(d-1)\times 48} - MDPseason_h)$, and is denoted by $MINVPseason_h$. However, the net minimum variance in wind power output during the $h$th half-hour of any day of the normative season is,

$$NETMINVseason_h = MINVPseason_h + MINVNPseason_{h+(d-1)\times 48} \quad (6)$$

where $d$ is the day number for which $MINVPseason_h$ is obtained. Similarly, the maximum variance in wind power output during the $h$th half-hour of any day of the normative season is the maximum value of $(NPseason_{h+(d-1)\times 48} - MDPseason_h)$, and is denoted by $MAXVPseason_h$. However, the net maximum variance in wind power output during the $h$th half-hour of any day of the normative season is,

$$NETMAXVseason_h = MAXVPseason_h + MAXVNPseason_{h+(d-1)\times 48} \quad (7)$$

where $d$ is the day number for which $MAXVPseason_h$ is obtained. Equations (6) and (7) follow from the fact that for the two independent random variables, $NORMPseason_h$ and $NPseason_h$, the variance of their sum is the sum of their individual variances [14].

The wind power outputs obtained for normative days of the four seasons are shown in Fig. 1. The solid lines represent the average wind power output during a normative day of the respective season. The dashed and dotted lines represent the maximum and minimum variation in normalized wind power output, respectively. The data indicates that there is more variation in wind power output on summer and spring normative days than in winter and fall normative days. This shows that there is a higher possibility of voltage violations occurring in the summer-spring months than in the winter-fall months. The actual variations in average wind power output during a normative day of 48 half-hours for different seasons are shown in Fig. 2. The average actual wind power output is observed to be the highest during the summer normative day and lowest during the winter normative day.

## C. Seasonal load modeling

Seasonal daily load models for residential customers were obtained using the methodology proposed in [15]. The daily load models for the summer season for commercial and industrial customers were obtained from [16]. In the load models, the daily load for any half-hour was represented in terms of the average daily peak load. For the $h$th half-hour of the $s$th season, let the normalized residential daily load be represented as $RNl_{s,h}$, where $s$ is 1, 2, 3 and 4, for the summer, spring, winter and fall seasons, respectively. For the $h$th half-hour of the summer season, let the normalized commercial and industrial daily load be represented as $CNl_{1,h}$ and $INl_{1,h}$, respectively. For the $h$th half-hour of any season, the residential daily load in MW can be obtained as shown in (8).

$$Rl_{s,h} = RNl_{s,h} \times \frac{RAe_s}{RDLC_s} \tag{8}$$

For the $h$th half-hour of the summer season, the commercial and industrial daily loads in MW can be obtained as,

$$Cl_{1,h} = CNl_{1,h} \times \frac{CAe_1}{CDLC_1} \tag{9}$$

$$Il_{1,h} = INl_{1,h} \times \frac{IAe_1}{IDLC_1} \tag{10}$$

where $RAe_s$ represents the seasonal average daily energy consumption for residential customers in MWh; $CAe_1$ and $IAe_1$ represent the summer average daily energy consumption for commercial and industrial customers, respectively, in MWh. The energy consumption values are obtained from [10]. In (8), $RDLC_s$ represents the area under the daily load curve for residential loads in the $s$th season. In (9) and (10), $CDLC_1$ and $IDLC_1$ represent the area under the daily load curves for commercial and industrial loads, respectively, in summer.

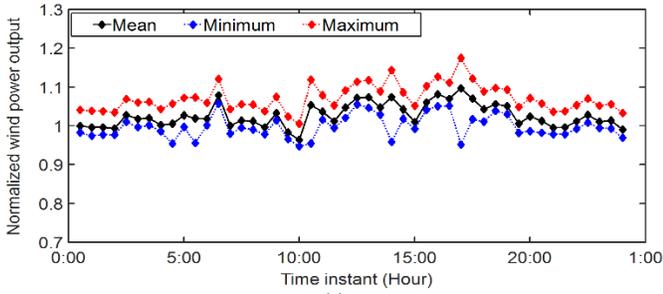

(a)

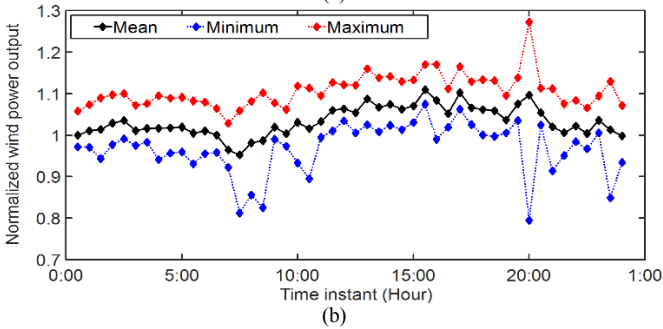

(b)

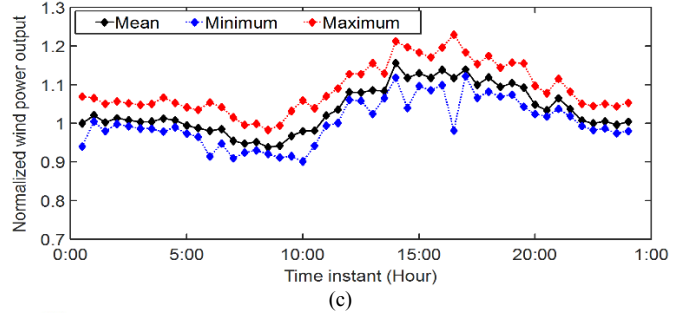

(c)

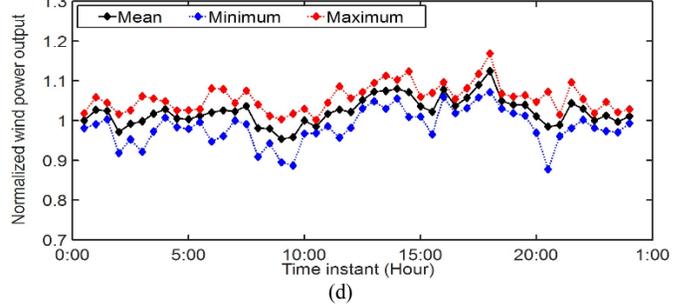

(d)

Fig. 1. Normative day wind power outputs for different seasons built using normalized 2007 to 2011 BPA wind power output data: (a) winter, (b) spring, (c) summer, (d) fall.

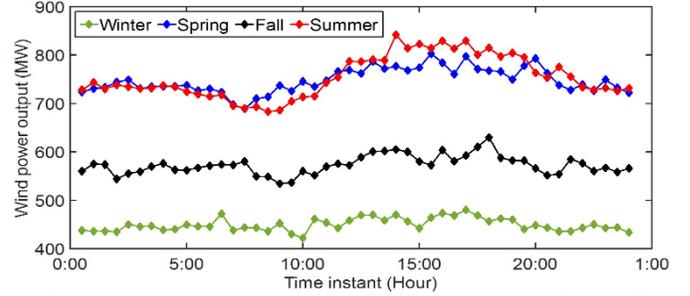

Fig. 2. Actual average wind power outputs for different seasons built using 2007 to 2011 BPA wind power data.

For the $h$th half-hour of the spring, fall or winter seasons, the commercial and industrial daily loads in MW can now be obtained as,

$$Cl_{s,h} = Cl_{1,h} \times \left(\frac{1 + LR_s}{1 + LR_1}\right) \tag{11}$$

$$Il_{s,h} = Il_{1,h} \times \left(\frac{1 + LR_s}{1 + LR_1}\right) \tag{12}$$

where $LR_s$ represents the ratio between the total passive and total active loads present in the system during the spring, winter or fall season; $LR_1$ represents the ratio between the total passive and total active loads present in the system during the summer season. These load ratios were evaluated using the technique described in [15]. The assumption in (11) and (12) is that the ratio between total passive and total active loads in the system for a season is the same for residential, commercial and industrial load types. Finally, for the $h$th half-hour of a season, the total system daily load as a percentage of the seasonal daily peak load can be obtained as,

$$Load_{s,h} = \frac{Rl_{s,h} + Cl_{s,h} + Il_{s,h}}{Pl_s} \times 100 \tag{13}$$

where $Pl_s$ represents the peak load in MW during a normative day of the $s$th season. The normative daily load demands obtained for the different seasons are shown in Fig. 3. It is observed that the peak normative loads for any day of a season occur between 10 AM and 5 PM. The actual variations in system load during a normative day for the four seasons are shown in Fig. 4. The daily system load is observed to be the highest during the winter normative day and lowest during the summer normative day.

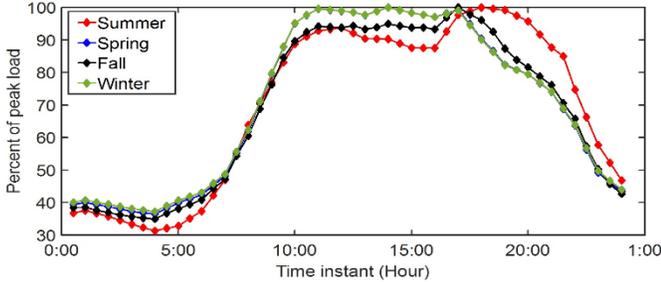

Fig. 3. System demand as percentage of peak load for different seasons built using 2007 to 2011 Oregon demand data.

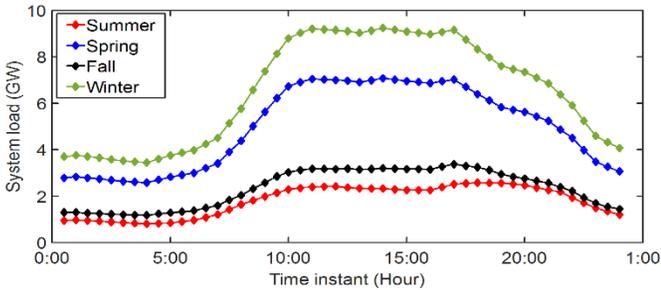

Fig. 4. Actual system demand for different seasons built using 2007 to 2011 Oregon demand data.

### III. MODEL VALIDATION FOR WIND GENERATION

To test the performance of the proposed measurement-based approach for wind power modeling, 30-minute wind power output data for the years 2012 to 2015 was obtained from the BPA website [9]. For each day of the years 2012 to 2015, the wind power outputs are normalized, as described in Section II-B. For any season of the years 2012 to 2015, if the half-hourly wind power output during a day exceeds ($NORMPseason_h + NETMAXVseason_h$) or is less than ($NORMPseason_h - NETMINVseason_h$), then it is identified as an outlier. The outliers for each year of a season are combined to obtain the total number of outliers for that season. The percentage of outliers for all four seasons is shown in Table I. It can be observed that the outliers for all seasons were relatively low (less than 8%), which validates the accuracy of the proposed approach. The outliers obtained during model testing can occur due to unforeseen weather conditions such as storms and hurricanes as well as due to presence of more noise in the data during periods of high telecommunication activities.

TABLE I. WIND MODEL TESTING RESULTS

| Season | Percentage of outliers |
|---|---|
| Winter | 7.8530 |
| Spring | 7.8583 |
| Summer | 6.0024 |
| Fall | 6.0963 |

### IV. VOLTAGE VIOLATION IDENTIFICATION

The MATPOWER toolbox [17] of MATLAB was used for performing the simulations. The IEEE 118-bus system was used as the test system for this analysis. For each season, the daily load profile was created based on the load modeling approach described in Section II-C; the seasonal load models are shown in Fig. 3 and Fig. 4. In the first step, for each season, an ACOPF was carried out for every half-hour of a normative day. The active and reactive loads at the system buses during the 48 half-hour periods were varied according to the normalized daily load models shown in Fig. 3. In this manner, 48 × 4 ACOPF result files were obtained. For any season, the optimal active and reactive power injections at the generator buses for each half-hourly period are obtained. These power injection values are utilized in the subsequent AC Power Flow (ACPF) studies to analyze the impacts of seasonal variations in wind generation on bus voltage violations.

As the wind power generation locations in the network are unknown, for the ACPF studies, the generator buses to which wind farms are connected are randomly selected such that the total wind penetration is approximately 50% of total generation. For each half-hourly normative period of a season, the active and reactive loads at the system buses change according to the corresponding daily load model show in Fig. 3. The power injections at the generator buses during any of the seasonal normative periods change according to (14) and (15),

$$POseason_{g,h} = WSseason_{g,h} \times PDseason_{g,h} \quad (14)$$
$$QOseason_{g,h} = WSseason_{g,h} \times QDseason_{g,h} \quad (15)$$

where $g = \{1,...,54\}$ and $h = \{1,...,48\}$. In (14) and (15), $PDseason_{g,h}$ and $QDseason_{g,h}$ represent the optimal active and reactive power injections at the $g$th generator bus for the $h$th normative period of a day of the season, respectively; $POseason_{g,h}$ and $QOseason_{g,h}$ represent the active and reactive power injections at the $g$th generator bus for any of the $h$ normative periods of a day of a season. For any period $h$, if a generator bus $g$ has not been selected as a wind generator bus, $WSseason_{g,h} = 1$. If a generator bus $g$ has been randomly selected as a bus with a wind generator connected to it, then for the season-independent approach,

$$WSseason_{g,h} = Pn_h \quad (16)$$

where $Pn_h$ is obtained from (3), with $V_{ci}$ = 3 m/s, $V_r$ = 12.5 m/s, and $V_{co}$ = 25 m/s [18]. If the same generator bus $g$ has been randomly selected as a bus with a wind generator connected to it, then for the season-focused approach,

$$WSseason_{g,h} = \frac{NORMPseason_h \times MPseason}{MPannual_h} \quad (17)$$

where $NORMPseason_h$ is obtained from Section II-B, $MPseason$ is the average BPA wind power output in MW for a season across the years 2007 to 2011 and $MPannual_h$ is the average value of the numerator on the right-hand side of (17) across the four seasons.

If the season-independent approach is considered, for $S$ different random selections of wind generator buses, 48 × $S$ ACPF result files are generated for the entire year (considering mean variation in wind power output during a year, and the annual daily load model). If the season-focused approach is considered, for $S$ different random selections of wind generator buses, 48 × $S$ ACPF result files are generated for each season, considering the mean variation in wind power output during a season and the corresponding seasonal daily load model. For the simulations carried out in this paper, $S = 100$.

For the season-focused approach $48 \times S$ ACPF result files can also be generated for each season, considering the minimum and maximum variations in wind power output during a season, and corresponding seasonal daily load models. If maximum variation in wind power output during a season is considered,

$$WSseason_{g,h} = \frac{(NORMPseason_h + NETMAXVseason_h) \times MXWPseason}{MXWPannual_h}$$
(18)

where $NETMAXVseason_h$ is obtained from (7), $MXWPseason$ is the maximum BPA wind power output in MW for a season across the years 2007 to 2011 and $MXWPannual_h$ is the maximum value of the numerator on the right-hand side of (18) across the four seasons. If minimum variation in wind power output during a season is considered,

$$WSseason_{g,h} = \frac{(NORMPseason_h - NETMINVseason_h) \times MNWPseason}{MNWPannual_h}$$
(19)

where $NETMINVseason_h$ is obtained from (6), $MNWPseason$ is the minimum BPA wind power output in MW for a season across the years 2007 to 2011 and $MNWPannual_h$ is the minimum value of the numerator on the right-hand side of (19) across the four seasons.

The performance comparison between season-independent and the season-focused approaches is shown in Table II. For the creation of this table, a voltage violation is assumed to occur if any bus voltage exceeds $\pm 5\%$ of its base case value. Table II only shows those buses that exceeded this range. The results shown in columns 2 to 5 consist of the seasonal voltage violations considering only the mean variation in wind power output during a season, and the corresponding seasonal load model. Each number in columns 2 to 6 shows the number of cases when voltage violations occur out of 4800 cases.

It is seen that by using the season-focused approach, the voltage violations at every bus for each season can be obtained, which is impossible if a season-independent approach is used. It is also seen that there are certain buses such as 90, 91, 92 that show voltage violations during two seasons and no voltage violations during the other seasons; this information is not revealed when the season-independent modeling approach is used. Finally, there are buses such as 101, 105 and 110 which do not show any voltage violations when the season-independent modeling approach is used, but show violations during spring and summer when the season-focused approach is employed. Therefore, by using the proposed modeling approach, one can identify locations that are vulnerable to variations in wind generation and load during different seasons of the year. This information can help in developing efficient optimization strategies such as determining optimal locations for placing portable and/or permanent energy storage units.

The voltage violations that might occur when the maximum and the minimum variations in wind power output, and the load variations for all seasons are considered is shown in Fig. 5. For the creation of this figure, a voltage violation is assumed to occur if any bus voltage exceeds the p.u. range [0.94,1.06].

Like Table II, only the buses where voltage violations occur are shown. The vertical axis shows the number of voltage violations (in percentage) that occur annually at different buses out of 38,400 cases ($= 4800 \times 2 \times 4$). Fig. 5 identifies the already stressed buses of the network that are most affected by the volatility in seasonal wind generation.

TABLE II. NUMBER OF VOLTAGE VIOLATIONS (OUT OF 4800) USING SEASON-INDEPENDENT AND SEASON-FOCUSED MODELING APPROACHES

| Bus # | Winter | Spring | Summer | Fall | Annual (season-independent) |
|---|---|---|---|---|---|
| 2 | 225 | 2710 | 3131 | 1326 | 2735 |
| 8 | 223 | 2520 | 2972 | 1316 | 2625 |
| 11 | 215 | 2306 | 2832 | 1293 | 2522 |
| 26 | 213 | 2233 | 2665 | 1226 | 2129 |
| 60 | 202 | 2160 | 2527 | 1049 | 2098 |
| 61 | 188 | 2141 | 2274 | 1042 | 1930 |
| 57 | 207 | 2106 | 2240 | 987 | 1876 |
| 62 | 193 | 2082 | 2202 | 980 | 1832 |
| 64 | 167 | 2068 | 2180 | 965 | 1748 |
| 68 | 160 | 2040 | 2152 | 940 | 1732 |
| 70 | 158 | 2028 | 2129 | 906 | 1715 |
| 71 | 135 | 2023 | 2084 | 753 | 1688 |
| 111 | 123 | 1991 | 2028 | 735 | 1624 |
| 117 | 108 | 1970 | 1998 | 661 | 1538 |
| 72 | 86 | 1925 | 1874 | 527 | 1421 |
| 22 | 72 | 1880 | 1805 | 316 | 1342 |
| 49 | 128 | 1485 | 1608 | 260 | 1236 |
| 25 | 145 | 1483 | 1607 | 234 | 1209 |
| 115 | 82 | 1015 | 1033 | 194 | 1174 |
| 114 | 71 | 549 | 562 | 217 | 642 |
| 80 | 13 | 481 | 518 | 23 | 537 |
| 83 | 14 | 456 | 515 | 18 | 512 |
| 47 | 11 | 425 | 483 | 15 | 420 |
| 48 | 0 | 393 | 462 | 16 | 328 |
| 90 | 0 | 379 | 448 | 13 | 287 |
| 91 | 0 | 346 | 340 | 0 | 268 |
| 92 | 0 | 347 | 321 | 0 | 149 |
| 100 | 0 | 341 | 296 | 0 | 72 |
| 101 | 0 | 129 | 267 | 0 | 0 |
| 105 | 0 | 76 | 254 | 0 | 0 |
| 110 | 0 | 70 | 217 | 0 | 0 |

The most vulnerable buses are determined by using the following index:

$$WV_b = \alpha_{1,b}/PV_{b,Case\ 1} + \alpha_{2,b}/PV_{b,Case\ 2} \quad (20)$$

where $WV_b$ refers to the weighted number of voltage violations at bus $b$; $\alpha_{1,b}$ and $\alpha_{2,b}$ refer to the ranking of the $b$th bus, in Table II and Fig. 5, respectively, on the basis of how vulnerable it is to seasonal voltage violations; $PV_{b,Case\ 1}$ refers to the fraction of the $4800 \times 4$ total cases for bus $b$ in Table II, which resulted in voltage violations; $PV_{b,Case\ 2}$ refers to the fraction of the 38,400 total cases for bus $b$ in Fig. 5, which resulted in voltage violations. The buses that were common to both Table II and Fig. 5 were identified as the set of most vulnerable buses, and arranged in increasing order of their $WV_b$ values. This was followed by buses which belonged to either Table II or Fig. 5, with the lower value of $WV_b$ indicating higher susceptibility. Based on (20), the 10 buses most vulnerable to seasonal voltage violations are 2, 49, 25, 115, 114, 80, 105, 76, 8 and 53. A flowchart describing the season-focused measurement-based approach is shown in Fig. 6.

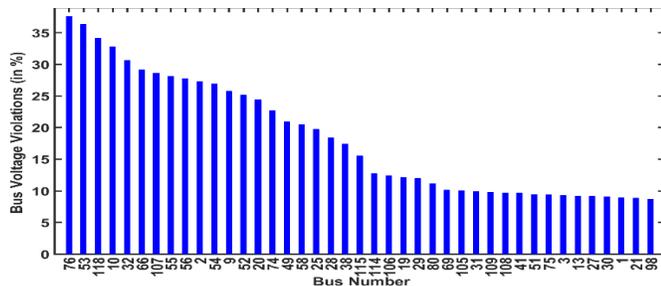

Fig. 5. Total annual bus voltage violations (in percentage) using the season-focused modeling approach.

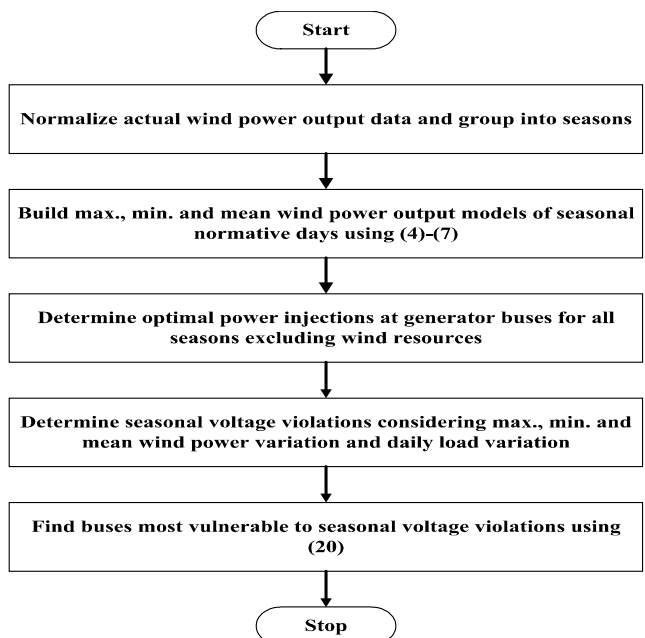

Fig. 6. Flowchart of the proposed seasonal bus voltage violation identification technique.

## V. CONCLUSIONS AND FUTURE SCOPE

This paper presents a measurement-based approach to wind power modeling that builds seasonal normative day wind power output models from actual wind power data. The model showed more than 92% testing accuracy when validated against several years of new wind power data. A load model that considered seasonal variations in its three components (residential, commercial, and industrial) was also developed. The proposed season-focused modeling approach was found to more effectively analyze the impacts of *seasonal* variations in wind power output and system load on the voltage profiles of the test system (IEEE 118-bus system) than a season-independent modeling approach.

The proposed season-focused approach has many future applications. It can be used in identifying locations where portable and/or permanent energy storage units can be placed. The proposed technique can also help in reducing the search space and computational time while solving non-linear and non-convex optimal energy storage placement problems for real power systems. It can also help determine how reserve resources can be more effectively used to minimize the *seasonal* voltage violations. Through discussions with their industry collaborators, the authors have found that power utilities are actively looking for such solutions. The final objective would be to combine seasonal variation in solar generation with that of load and wind to obtain a realistic comprehensive visualization of the impacts that DERs can have on the electric grid.

## VI. REFERENCES

[1] Advantages and Challenges of Wind Energy [Online]. Available: http://energy.gov/eere/wind/advantages-and-challenges-wind-energy
[2] J. Dowds et al., "A review of large-scale wind integration studies," *Renew. Sust. Energ. Rev.*, vol. 49, pp. 768–794, Sept. 2015.
[3] M. A. Chowdhury, W. Shen, N. Hosseinzadeh, and H. R. Pota, "Transient stability of power system integrated with doubly fed induction generator wind farms," *IET Renew. Power Gen.*, vol. 9, no. 2, pp. 184–194, Feb. 2015.
[4] Y. Pan et al., "Admissible region of large-scale uncertain wind generation considering small-signal stability of power systems," *IEEE Trans. Sustain. Energy*, vol. 7, no. 4, pp. 1611–1623, Oct. 2016.
[5] M. Ghofrani, A. Arabali, M. Etezadi-Amoli, and M. S. Fadali, "A framework for optimal placement of energy storage units within a power system with high wind energy penetration," *IEEE Trans. Sustain. Energy*, vol. 4, no. 2, pp. 443–442, Apr. 2013.
[6] S. Wen, H. Lan, Q. Fu, D. C. Yu, and L. Zhang, "Economic allocation for energy storage system considering wind power distribution," *IEEE Trans. Power Syst.*, vol. 30, no.2, pp. 644–652, Mar. 2015.
[7] Y. Ge, A. J. Flueck, D. K. Kim, J. B. Ahn, J. D. Lee, and D. Y. Kwon, "An event-oriented method for online load modeling based on synchrophasor data," *IEEE Trans. Smart Grid*, vol. 6, no. 4, pp. 2060-2068, Jul. 2015.
[8] H. Dong, H. Renmu, X. Yanhui, M. Jin, and H. Mei, "Measurement-based load modeling validation by artificial three-phase short circuit tests in north east power grid," in *Proc. IEEE Power Eng. Soc. General Meeting*, Tampa, FL, pp. 1-6, 24-28 Jun. 2007.
[9] Wind History (BPA) [Online]. Available: http://transmission.bpa.gov/business/operations/wind/
[10] Electric Power Monthly - Electricity (U.S. EIA) [Online]. Available: https://www.eia.gov/electricity/monthly/
[11] J. Ma, Y. Qiu, Y. Li, and J. S. Thorp, "Stability analysis of power system with multiple operating conditions considering the stochastic characteristic of wind speed," *IET Gener. Transm. Distrib.*, vol. 10, no. 4, pp. 1056–1066, Mar. 2016.
[12] C. Sun, Z. Bie, M. Xie, and G. Ning, "Effects of wind speed probabilistic and possibilistic uncertainties on generation system adequacy," *IET Gener. Transm. Distrib.*, vol. 9, no. 4, pp. 339–347, Mar. 2015.
[13] Meteorological Information from BPA Weather Sites [Online]. Available: https://transmission.bpa.gov/business/operations/wind/MetData/MetMonthly.aspx?id=Troutdale&name=Troutdale.
[14] Mean and Variance of Random Variables [Online]. Available: http://www.stat.yale.edu/Courses/1997-98/101/rvmnvar.htm.
[15] R. Subbiah, A. Pal, E. K. Nordberg, A. Marathe, and M. V. Marathe, "Energy Demand Model for Residential Sector: A First Principles Approach," *IEEE Trans. Sustain. Energy*, vol. PP, no. 99, pp. 1–10, Feb. 2017.
[16] J. C. Cepeda, J. L. Rueda, D. G. Colome, and D. E. Echeverria, "Real-time transient stability assessment based on center-of-inertia estimation from phasor measurement unit records," *IET Gener. Transm. Distrib.*, vol. 8, no. 8, pp. 1363–1376, Aug. 2014.
[17] MATPOWER (Oct. 20, 2016). "A MATLAB Power System Simulation Package" [Online]. Available: http://www.pserc.cornell.edu/matpower/.
[18] M. Aien, M. F. Firuzabad, and F. Aminifar, "Unscented transformation-based probabilistic optimal power flow for modeling the effect of wind power generation," *Turk J. Elec. Eng. Comp. Sci.*, vol. 21, no.5, pp. 1284–1301, Aug. 2013.